
\magnification=\magstephalf 
\newbox\SlashedBox 
\def\slashed#1{\setbox\SlashedBox=\hbox{#1}
\hbox to 0pt{\hbox to 1\wd\SlashedBox{\hfil/\hfil}\hss}{#1}}
\def\hboxtosizeof#1#2{\setbox\SlashedBox=\hbox{#1}
\hbox to 1\wd\SlashedBox{#2}}

\def\mathslashed#1{\setbox\SlashedBox=\hbox{$#1$}
\hbox to 0pt{\hbox to 1\wd\SlashedBox{\hfil/\hfil}\hss}#1}

\def\ifsmall{\iffalse}  
\def\titlepagefont{}  

\def\DefineTeXgraphics{%
\special{ps::[global] /TeXgraphics { } def}}  

\def\today{\ifcase\month\or January\or February\or March\or April\or May
\or June\or July\or August\or September\or October\or November\or
December\fi\space\number\day, \number\year}
\def\eatPrefix19{}
\def\Year{\expandafter\eatPrefix\the\year}
\newcount\hours \newcount\minutes
\def\monthname{\ifcase\month\or
January\or February\or March\or April\or May\or June\or July\or
August\or September\or October\or November\or December\fi}
\def\shortmonthname{\ifcase\month\or
Jan\or Feb\or Mar\or Apr\or May\or Jun\or Jul\or
Aug\or Sep\or Oct\or Nov\or Dec\fi}

\def\TimeStamp{\hours\the\time\divide\hours by60%
\minutes -\the\time\divide\minutes by60\multiply\minutes by60%
\advance\minutes by\the\time%
${\rm \shortmonthname}\cdot\if\day<10{}0\fi\the\day\cdot\the\year%
\qquad\the\hours:\if\minutes<10{}0\fi\the\minutes$}




\def\Title#1{%
\vskip 1in{\titlefont\centerline{#1}}\vskip .5in}
 


\newif\ifdraftmode
\newif\ifleftlabels  

\def\nolabels{\def\wrlabeL##1{}\def\eqlabeL##1{}\def\reflabeL##1{}}
\def\writelabels{\def\wrlabeL##1{\leavevmode\vadjust{\rlap{\smash%
{\line{{\escapechar=` \hfill\rlap{\sevenrm\hskip.03in\string##1}}}}}}}%
\def\eqlabeL##1{{\escapechar-1\rlap{\sevenrm\hskip.05in\string##1}}}%
\def\reflabeL##1{\noexpand\rlap{\noexpand\sevenrm[\string##1]}}}
\def\writeleftlabels{\def\wrlabeL##1{\leavevmode\vadjust{\rlap{\smash%
{\line{{\escapechar=` \hfill\rlap{\sevenrm\hskip.03in\string##1}}}}}}}%
\def\eqlabeL##1{{\escapechar-1%
\rlap{\sixrm\hskip.05in\string##1}%
\llap{\sevenrm\string##1\hskip.03in\hbox to \hsize{}}}}%
\def\reflabeL##1{\noexpand\rlap{\noexpand\sevenrm[\string##1]}}}
\nolabels

\newdimen\fullhsize
\newdimen\hstitle
\hstitle=\hsize 
\newdimen\hsbody
\hsbody=\hsize 
\newdimen\hbodyoffset
\hbodyoffset=\hoffset 
\newbox\leftpage
\def\abstract#1{#1}
\def\rotated{\special{ps: landscape}
\magnification=1000  
\baselineskip=14pt
\global\hstitle=9truein\global\hsbody=4.75truein
\global\vsize=7truein\global\voffset=-.31truein
\global\hoffset=-0.54in\global\hbodyoffset=-.54truein
\global\fullhsize=10truein
\def\DefineTeXgraphics{%
\special{ps::[global] 
/TeXgraphics {currentpoint translate 0.7 0.7 scale
              -80 0.72 mul -1000 0.72 mul translate} def}}
\let\lr=L
\def\ifsmall{\iftrue}
\def\titlepagefont{\twelvepoint}
\trueseventeenpoint
\def\almostshipout##1{\if L\lr \count1=1
      \global\setbox\leftpage=##1 \global\let\lr=R
   \else \count1=2
      \shipout\vbox{\hbox to\fullhsize{\box\leftpage\hfil##1}}
      \global\let\lr=L\fi}

\output={\ifnum\count0=1 
 \shipout\vbox{\hbox to \fullhsize{\hfill\pagebody\hfill}}\advancepageno
 \else
 \almostshipout{\leftline{\vbox{\pagebody\makefootline}}}\advancepageno 
 \fi}

\def\abstract##1{{\leftskip=1.5in\rightskip=1.5in ##1\par}} }

\def\linemessage#1{\immediate\write16{#1}}

\global\newcount\secno \global\secno=0
\global\newcount\appno \global\appno=0
\global\newcount\meqno \global\meqno=1
\global\newcount\subsecno \global\subsecno=0
\global\newcount\figno \global\figno=0

\newif\ifAnyCounterChanged
\let\terminator=\relax
\def\normalize#1{\ifx#1\terminator\let\next=\relax\else%
\if#1i\aftergroup i\else\if#1v\aftergroup v\else\if#1x\aftergroup x%
\else\if#1l\aftergroup l\else\if#1c\aftergroup c\else%
\if#1m\aftergroup m\else%
\if#1I\aftergroup I\else\if#1V\aftergroup V\else\if#1X\aftergroup X%
\else\if#1L\aftergroup L\else\if#1C\aftergroup C\else%
\if#1M\aftergroup M\else\aftergroup#1\fi\fi\fi\fi\fi\fi\fi\fi\fi\fi\fi\fi%
\let\next=\normalize\fi%
\next}
\def\makeNormal#1#2{\def\doNormalDef{\edef#1}\begingroup%
\aftergroup\doNormalDef\aftergroup{\normalize#2\terminator\aftergroup}%
\endgroup}

\def\warnIfChanged#1#2{%
\ifundef#1
\else\begingroup%
\edef\oldDefinitionOfCounter{#1}\edef\newDefinitionOfCounter{#2}%
\ifx\oldDefinitionOfCounter\newDefinitionOfCounter%
\else%
\linemessage{Warning: definition of \noexpand#1 has changed.}%
\global\AnyCounterChangedtrue\fi\endgroup\fi}

\def\Section#1{\global\advance\secno by1\relax\global\meqno=1%
\global\subsecno=0%
\bigbreak\bigskip
\centerline{\twelvepoint \bf %
\the\secno. #1}%
\par\nobreak\medskip\nobreak}
\def\tagsection#1{%
\warnIfChanged#1{\the\secno}%
\xdef#1{\the\secno}%
\ifWritingAuxFile\immediate\write\auxfile{\noexpand\xdef\noexpand#1{#1}}\fi%
}
\def\section{\Section}
\def\Subsection#1{\global\advance\subsecno by1\relax\medskip %
\leftline{\bf\the\secno.\the\subsecno\ #1}%
\par\nobreak\smallskip\nobreak}
\def\tagsubsection#1{%
\warnIfChanged#1{\the\secno.\the\subsecno}%
\xdef#1{\the\secno.\the\subsecno}%
\ifWritingAuxFile\immediate\write\auxfile{\noexpand\xdef\noexpand#1{#1}}\fi%
}

\def\subsection{\Subsection}

\def\romappno{\uppercase\expandafter{\romannumeral\appno}}
\def\makeNormalizedRomappno{%
\expandafter\makeNormal\expandafter\normalizedromappno%
\expandafter{\romannumeral\appno}%
\edef\normalizedromappno{\uppercase{\normalizedromappno}}}
\def\Appendix#1{\global\advance\appno by1\relax\global\meqno=1\global\secno=0%
\global\subsecno=0%
\bigbreak\bigskip
\centerline{\twelvepoint \bf Appendix %
\romappno. #1}%
\par\nobreak\medskip\nobreak}
\def\tagappendix#1{\makeNormalizedRomappno%
\warnIfChanged#1{\normalizedromappno}%
\xdef#1{\normalizedromappno}%
\ifWritingAuxFile\immediate\write\auxfile{\noexpand\xdef\noexpand#1{#1}}\fi%
}
\def\appendix{\Appendix}
\def\Subappendix#1{\global\advance\subsecno by1\relax\medskip %
\leftline{\bf\romappno.\the\subsecno\ #1}%
\par\nobreak\smallskip\nobreak}
\def\tagsubappendix#1{\makeNormalizedRomappno%
\warnIfChanged#1{\normalizedromappno.\the\subsecno}%
\xdef#1{\normalizedromappno.\the\subsecno}%
\ifWritingAuxFile\immediate\write\auxfile{\noexpand\xdef\noexpand#1{#1}}\fi%
}

\def\eqn#1{\makeNormalizedRomappno%
\ifnum\secno>0%
  \warnIfChanged#1{\the\secno.\the\meqno}%
  \eqno(\the\secno.\the\meqno)\xdef#1{\the\secno.\the\meqno}%
     \global\advance\meqno by1
\else\ifnum\appno>0%
  \warnIfChanged#1{\normalizedromappno.\the\meqno}%
  \eqno({\rm\romappno}.\the\meqno)%
      \xdef#1{\normalizedromappno.\the\meqno}%
     \global\advance\meqno by1
\else%
  \warnIfChanged#1{\the\meqno}%
  \eqno(\the\meqno)\xdef#1{\the\meqno}%
     \global\advance\meqno by1
\fi\fi%
\eqlabeL#1%
\ifWritingAuxFile\immediate\write\auxfile{\noexpand\xdef\noexpand#1{#1}}\fi%
}
\def\defeqn#1{\makeNormalizedRomappno%
\ifnum\secno>0%
  \warnIfChanged#1{\the\secno.\the\meqno}%
  \xdef#1{\the\secno.\the\meqno}%
     \global\advance\meqno by1
\else\ifnum\appno>0%
  \warnIfChanged#1{\normalizedromappno.\the\meqno}%
  \xdef#1{\normalizedromappno.\the\meqno}%
     \global\advance\meqno by1
\else%
  \warnIfChanged#1{\the\meqno}%
  \xdef#1{\the\meqno}%
     \global\advance\meqno by1
\fi\fi%
\eqlabeL#1%
\ifWritingAuxFile\immediate\write\auxfile{\noexpand\xdef\noexpand#1{#1}}\fi%
}
\def\anoneqn{\makeNormalizedRomappno%
\ifnum\secno>0
  \eqno(\the\secno.\the\meqno)%
     \global\advance\meqno by1
\else\ifnum\appno>0
  \eqno({\rm\normalizedromappno}.\the\meqno)%
     \global\advance\meqno by1
\else
  \eqno(\the\meqno)%
     \global\advance\meqno by1
\fi\fi%
}
\def\mfig#1#2{\ifx#20
\else\global\advance\figno by1%
\relax#1\the\figno%
\warnIfChanged#2{\the\figno}%
\xdef#2{\the\figno}%
\reflabeL#2%
\ifWritingAuxFile\immediate\write\auxfile{\noexpand\xdef\noexpand#2{#2}}\fi\fi%
}

\catcode`@=11 

\newif\ifFiguresInText\FiguresInTexttrue
\newif\if@FigureFileCreated
\newwrite\capfile
\newwrite\figfile

\newif\ifcaption
\captiontrue
\def\captionsize{\tenrm}
\def\PlaceTextFigure#1#2#3#4{%
\vskip 0.5truein%
#3\hfil\epsfbox{#4}\hfil\break%
\ifcaption\hfil\vbox{\captionsize Figure #1. #2}\hfil\fi%
\vskip10pt}
\def\PlaceEndFigure#1#2{%
\epsfxsize=\hsize\epsfbox{#2}\vfill\centerline{Figure #1.}\eject}

\def\LoadFigure#1#2#3#4{%
\ifundef#1{\phantom{\mfig{}#1}}\else
\fi%
\ifFiguresInText
\PlaceTextFigure{#1}{#2}{#3}{#4}%
\else
\if@FigureFileCreated\else%
\immediate\openout\capfile=\jobname.caps%
\immediate\openout\figfile=\jobname.figs%
@FigureFileCreatedtrue\fi%
\immediate\write\capfile{\noexpand\item{Figure \noexpand#1.\ }{#2}\vskip10pt}%
\immediate\write\figfile{\noexpand\PlaceEndFigure\noexpand#1{\noexpand#4}}%
\fi}

\def\listfigs{\ifFiguresInText\else%
\vfill\eject\immediate\closeout\capfile
\immediate\closeout\figfile%
\centerline{{\bf Figures}}\bigskip\frenchspacing%
\catcode`@=11 
\def\captionsize{\tenrm}
\input \jobname.caps\vfill\eject\nonfrenchspacing%
\catcode`\@=\active
\catcode`@=12  
\input\jobname.figs\fi}

\font\ninerm=cmr9
\font\eightrm=cmr8
\font\sixrm=cmr6

\def\loadtrueseventeenpoint{
 \font\seventeenrm=cmr10 at 17.28truept
 \font\seventeeni=cmmi10 at 17.28truept
 \font\seventeenbf=cmbx10 at 17.28truept
 \font\seventeenit=cmti10 at 17.28truept
 \font\seventeensl=cmsl10 at 17.28truept
 \font\seventeensy=cmsy10 at 17.28truept
}
\def\loadfourteenpoint{
\font\fourteenrm=cmr10 at 14.4pt
\font\fourteeni=cmmi10 at 14.4pt
\font\fourteenit=cmti10 at 14.4pt
\font\fourteensl=cmsl10 at 14.4pt
\font\fourteensy=cmsy10 at 14.4pt
\font\fourteenbf=cmbx10 at 14.4pt
}
\def\loadtruetwelvepoint{
\font\twelverm=cmr10 at 12truept
\font\twelvei=cmmi10 at 12truept
\font\twelveit=cmti10 at 12truept
\font\twelvesl=cmsl10 at 12truept
\font\twelvesy=cmsy10 at 12truept
\font\twelvebf=cmbx10 at 12truept
}

\font\ninei=cmmi9
\font\eighti=cmmi8
\font\sixi=cmmi6
\skewchar\ninei='177 \skewchar\eighti='177 \skewchar\sixi='177

\font\ninesy=cmsy9
\font\eightsy=cmsy8
\font\sixsy=cmsy6
\skewchar\ninesy='60 \skewchar\eightsy='60 \skewchar\sixsy='60

\font\ninebf=cmbx9
\font\eightbf=cmbx8
\font\sixbf=cmbx6

\font\ninett=cmtt9
\font\eighttt=cmtt8

\hyphenchar\tentt=-1 
\hyphenchar\ninett=-1
\hyphenchar\eighttt=-1         

\font\ninesl=cmsl9
\font\eightsl=cmsl8

\font\nineit=cmti9
\font\eightit=cmti8

                      
\newskip\ttglue
\def\tenpoint{\def\rm{\fam0\tenrm}%
  \textfont0=\tenrm \scriptfont0=\sevenrm \scriptscriptfont0=\fiverm
  \textfont1=\teni \scriptfont1=\seveni \scriptscriptfont1=\fivei
  \textfont2=\tensy \scriptfont2=\sevensy \scriptscriptfont2=\fivesy
  \textfont3=\tenex \scriptfont3=\tenex \scriptscriptfont3=\tenex
  \def\it{\fam\itfam\tenit}\textfont\itfam=\tenit
  \def\sl{\fam\slfam\tensl}\textfont\slfam=\tensl
  \def\bf{\fam\bffam\tenbf}\textfont\bffam=\tenbf \scriptfont\bffam=\sevenbf
  \scriptscriptfont\bffam=\fivebf
  \normalbaselineskip=12pt
  \let\sc=\eightrm
  \let\big=\tenbig
  \setbox\strutbox=\hbox{\vrule height8.5pt depth3.5pt width\z@}%
  \normalbaselines\rm}

\def\twelvepoint{\def\rm{\fam0\twelverm}%
  \textfont0=\twelverm \scriptfont0=\ninerm \scriptscriptfont0=\sevenrm
  \textfont1=\twelvei \scriptfont1=\ninei \scriptscriptfont1=\seveni
  \textfont2=\twelvesy \scriptfont2=\ninesy \scriptscriptfont2=\sevensy
  \textfont3=\tenex \scriptfont3=\tenex \scriptscriptfont3=\tenex
  \def\it{\fam\itfam\twelveit}\textfont\itfam=\twelveit
  \def\sl{\fam\slfam\twelvesl}\textfont\slfam=\twelvesl
  \def\bf{\fam\bffam\twelvebf}\textfont\bffam=\twelvebf%
  \scriptfont\bffam=\ninebf
  \scriptscriptfont\bffam=\sevenbf
  \normalbaselineskip=12pt
  \let\sc=\eightrm
  \let\big=\tenbig
  \setbox\strutbox=\hbox{\vrule height8.5pt depth3.5pt width\z@}%
  \normalbaselines\rm}

\def\fourteenpoint{\def\rm{\fam0\fourteenrm}%
  \textfont0=\fourteenrm \scriptfont0=\tenrm \scriptscriptfont0=\sevenrm
  \textfont1=\fourteeni \scriptfont1=\teni \scriptscriptfont1=\seveni
  \textfont2=\fourteensy \scriptfont2=\tensy \scriptscriptfont2=\sevensy
  \textfont3=\tenex \scriptfont3=\tenex \scriptscriptfont3=\tenex
  \def\it{\fam\itfam\fourteenit}\textfont\itfam=\fourteenit
  \def\sl{\fam\slfam\fourteensl}\textfont\slfam=\fourteensl
  \def\bf{\fam\bffam\fourteenbf}\textfont\bffam=\fourteenbf%
  \scriptfont\bffam=\tenbf
  \scriptscriptfont\bffam=\sevenbf
  \normalbaselineskip=17pt
  \let\sc=\elevenrm
  \let\big=\tenbig                                          
  \setbox\strutbox=\hbox{\vrule height8.5pt depth3.5pt width\z@}%
  \normalbaselines\rm}

\def\seventeenpoint{\def\rm{\fam0\seventeenrm}%
  \textfont0=\seventeenrm \scriptfont0=\fourteenrm \scriptscriptfont0=\tenrm
  \textfont1=\seventeeni \scriptfont1=\fourteeni \scriptscriptfont1=\teni
  \textfont2=\seventeensy \scriptfont2=\fourteensy \scriptscriptfont2=\tensy
  \textfont3=\tenex \scriptfont3=\tenex \scriptscriptfont3=\tenex
  \def\it{\fam\itfam\seventeenit}\textfont\itfam=\seventeenit
  \def\sl{\fam\slfam\seventeensl}\textfont\slfam=\seventeensl
  \def\bf{\fam\bffam\seventeenbf}\textfont\bffam=\seventeenbf%
  \scriptfont\bffam=\fourteenbf
  \scriptscriptfont\bffam=\twelvebf
  \normalbaselineskip=21pt
  \let\sc=\fourteenrm
  \let\big=\tenbig                                          
  \setbox\strutbox=\hbox{\vrule height 12pt depth 6pt width\z@}%
  \normalbaselines\rm}

\def\ninepoint{\def\rm{\fam0\ninerm}%
  \textfont0=\ninerm \scriptfont0=\sixrm \scriptscriptfont0=\fiverm
  \textfont1=\ninei \scriptfont1=\sixi \scriptscriptfont1=\fivei
  \textfont2=\ninesy \scriptfont2=\sixsy \scriptscriptfont2=\fivesy
  \textfont3=\tenex \scriptfont3=\tenex \scriptscriptfont3=\tenex
  \def\it{\fam\itfam\nineit}\textfont\itfam=\nineit
  \def\sl{\fam\slfam\ninesl}\textfont\slfam=\ninesl
  \def\bf{\fam\bffam\ninebf}\textfont\bffam=\ninebf \scriptfont\bffam=\sixbf
  \scriptscriptfont\bffam=\fivebf
  \normalbaselineskip=11pt
  \let\sc=\sevenrm
  \let\big=\ninebig
  \setbox\strutbox=\hbox{\vrule height8pt depth3pt width\z@}%
  \normalbaselines\rm}

\def\eightpoint{\def\rm{\fam0\eightrm}%
  \textfont0=\eightrm \scriptfont0=\sixrm \scriptscriptfont0=\fiverm%
  \textfont1=\eighti \scriptfont1=\sixi \scriptscriptfont1=\fivei%
  \textfont2=\eightsy \scriptfont2=\sixsy \scriptscriptfont2=\fivesy%
  \textfont3=\tenex \scriptfont3=\tenex \scriptscriptfont3=\tenex%
  \def\it{\fam\itfam\eightit}\textfont\itfam=\eightit%
  \def\sl{\fam\slfam\eightsl}\textfont\slfam=\eightsl%
  \def\bf{\fam\bffam\eightbf}\textfont\bffam=\eightbf \scriptfont\bffam=\sixbf%
  \scriptscriptfont\bffam=\fivebf%
  \normalbaselineskip=9pt%
  \let\sc=\sixrm%
  \let\big=\eightbig%
  \setbox\strutbox=\hbox{\vrule height7pt depth2pt width\z@}%
  \normalbaselines\rm}

\def\tenbig#1{{\hbox{$\left#1\vbox to8.5pt{}\right.\n@space$}}}
\def\ninebig#1{{\hbox{$\textfont0=\tenrm\textfont2=\tensy
  \left#1\vbox to7.25pt{}\right.\n@space$}}}
\def\eightbig#1{{\hbox{$\textfont0=\ninerm\textfont2=\ninesy
  \left#1\vbox to6.5pt{}\right.\n@space$}}}

\def\footnote#1{\edef\@sf{\spacefactor\the\spacefactor}#1\@sf
      \insert\footins\bgroup\eightpoint
      \interlinepenalty100 \let\par=\endgraf
        \leftskip=\z@skip \rightskip=\z@skip
        \splittopskip=10pt plus 1pt minus 1pt \floatingpenalty=20000
        \smallskip\item{#1}\bgroup\strut\aftergroup\@foot\let\next}
\skip\footins=12pt plus 2pt minus 4pt 
\dimen\footins=30pc 

\newinsert\margin
\dimen\margin=\maxdimen
\def\titlefont{\seventeenpoint}
\loadtruetwelvepoint 
\loadtrueseventeenpoint

\def\eatOne#1{}
\def\ifundef#1{\expandafter\ifx%
\csname\expandafter\eatOne\string#1\endcsname\relax}
\def\notTrue{\iffalse}\def\isTrue{\iftrue}
\def\ifdef#1{{\ifundef#1%
\aftergroup\notTrue\else\aftergroup\isTrue\fi}}
\def\use#1{\ifundef#1\linemessage{Warning: \string#1 is undefined.}%
{\tt \string#1}\else#1\fi}



%
\catcode`"=11
\let\quote="
\catcode`"=12
\chardef\foo="22
\global\newcount\refno \global\refno=1
\newwrite\rfile
\newlinechar=`\^^J
\def\@ref#1#2{\the\refno\n@ref#1{#2}}
\def\h@ref#1#2#3{\href{#3}{\the\refno}\n@ref#1{#2}}
\def\n@ref#1#2{\xdef#1{\the\refno}%
\ifnum\refno=1\immediate\openout\rfile=\jobname.refs\fi%
\immediate\write\rfile{\noexpand\item{[\noexpand#1]\ }#2.}%
\global\advance\refno by1}
\def\nref{\n@ref} 
\def\ref{\@ref}   
\def\hrref{\h@ref}
\def\lref#1#2{\the\refno\xdef#1{\the\refno}%
\ifnum\refno=1\immediate\openout\rfile=\jobname.refs\fi%
\immediate\write\rfile{\noexpand\item{[\noexpand#1]\ }#2\semi}%
\global\advance\refno by1}
\def\cref#1{\immediate\write\rfile{#1\semi}}

\def\preref#1#2{\gdef#1{\@ref#1{#2}}}

\def\semi{;\hfil\noexpand\break}

\def\listrefs{\vfill\eject\immediate\closeout\rfile
\centerline{{\bf References}}\bigskip\frenchspacing%
\input \jobname.refs\vfill\eject\nonfrenchspacing}

\def\inputAuxIfPresent#1{\immediate\openin1=#1
\ifeof1\message{No file \auxfileName; I'll create one.
}\else\closein1\relax\input\auxfileName\fi%
}




\newif\ifWritingAuxFile
\newwrite\auxfile
\def\SetUpAuxFile{%
\xdef\auxfileName{\jobname.aux}%
\inputAuxIfPresent{\auxfileName}%
\WritingAuxFiletrue%
\immediate\openout\auxfile=\auxfileName}

\def\L{\left(}\def\R{\right)}
\def\LP{\left.}\def\RP{\right.}
\def\LB{\left[}\def\RB{\right]}

\def\RV{\right|}

\def\bye{\par\vfill\supereject%
\ifAnyCounterChanged\linemessage{
Some counters have changed.  Re-run tex to fix them up.}\fi%
\end}

\catcode`\@=\active
\catcode`@=12  
\catcode`\"=\active




\hfuzz 30 pt
\SetUpAuxFile
 
 \loadfourteenpoint
\font\smc=cmcsc10

\noindent

\def\Re{\mathop{\rm Re}}
\def\Im{\mathop{\rm Im}}

\preref\DR{P. J. Davis and P. Rabinowitz, 
 {\it Methods of Numerical Integration\/}, 2nd Edition, Academic Press, 1984}
\preref\GR{M. Gl{\accent 127 u}ck and E. Reya, Phys.\ Rev.\ D14:3034 (1976)}
\preref\GRV{M. Gl{\accent 127 u}ck, E. Reya, and A. Vogt, Z. Phys.\ C48:471 (1990)}
\preref\MRS{A. D. Martin, R. G. Roberts, and W. J. Stirling,
Phys.\ Lett.\ B387:419 (1996) [hep-ph/9606345]; 
Phys.\ Lett.\ B354:155 (1995) [hep-ph/9502336];
Int.\ J.\ Mod.\ Phys.\ A10:2885 (1995)}
\preref\CTEQ{H. L. Lai, J. Huston, S. Kuhlmann, F. Olness, J. F. Owens, D. Soper, 
W. K. Tung, H. Weerts, Phys.\ Rev.\ D55:1280 (1997) [hep-ph/9606399]\semi
H. L. Lai, J. Botts, J. Huston, J. G. Morfin, J. F. Owens, J. W. Qiu, W. K. Tung, 
H. Weerts, Phys.\ Rev.\ D51:4763 (1995) [hep-ph/9410404]}
\preref\EvolEqnRef{G. Altarelli and G. Parisi, Nucl.\ Phys.\ B126:298 (1977)}
\preref\BetaRefs{D. J. Gross and F. W. Wilczek, Phys.\ Rev.\ Lett.\ 30:1343 (1973)\semi
                 H. D. Politzer, Phys.\ Rev.\ Lett.\ 30:1346 (1973)\semi
                 W. Caswell, Phys.\ Rev.\ Lett.\ 33:224 (1974)\semi
                 D. R. T. Jones, Nucl.\ Phys.\ B87:127 (1975)}
\preref\Blum{J. Bl{\accent 127 u}mlein, S. Riemersma, W. L. van Neerven, and A. Vogt,
             Nucl.\ Phys.\ Proc.\ Suppl.\ 51C:97 (1996)
             [hep-ph/9609217]}
\preref\Compare{J. Bl{\accent 127 u}mlein, S. Riemersma, M. Botje, C. Pascaud, F. Zomer,
                W. L. van Neerven, A. Vogt, in
                {\it Workshop on Future Physics 
                at HERA, Hamburg, Germany, 25-26 Sep 1995\/} [hep-ph/9609400]}
\preref\Mellin{M. Diemoz, F. Ferroni, E. Longo, and G. Martinelli, 
               Z. Phys.\ C39:21 (1988)}
\preref\Direct{M. Virchaux and A. Ouraou, DPhPE 87--15\semi
               M. Virchaux, PhD Thesis, University of Paris--7 (1988)\semi
               A. Ouraou, PhD Thesis, University of Paris--11 (1988)\semi
               M. Botje, {\smc qcdnum}15: {\it A fast QCD evolution program\/},
                 to appear\semi
               C. Pascaud and F. Zomer, HERA H1 note H1--11/94--404}
\preref\FP{W. Furmanski and R. Petronzio, Z. Phys.\ C11:293 (1982)}
\preref\FKL{E. G. Floratos, C. Kounnas, and R. Lacaze, 
Nucl. Phys. B192:417 (1981)}
\preref\GG{W. T.\ Giele and E. W. N.\ Glover,
Phys.\ Rev.\ D46:1980 (1992)}
\preref\GGK{W. T. Giele, E. W. N. Glover, and D. A. Kosower, Nucl. Phys. B403:633 (1993)
[hep-ph/9302225]}
\preref\FKS{S. Frixione, Z. Kunszt, and A. Signer, Nucl.\ Phys.\ B467:399 (1996) 
[hep-ph/9512328]}
\preref\CS{S. Catani and M. H. Seymour, Nucl.\ Phys.\ B485:291 (1997) [hep-ph/9605323]}
\preref\Graudenz{D. Graudenz, M. Hampel, A. Vogt, C. Berger, Z.\ Phys.\ C70:77 (1996) 
  [hep-ph/9506333]}
\preref\EfficientEvolution{D. A. Kosower, Saclay preprint SPhT--T97/043
[hep-ph/9706213]}
$\null$
 
\vskip -.6 cm
  
\nopagenumbers
\noindent hep-ph/9708392 \hfill Saclay--SPhT/T97--101

\vskip -2.0 cm 

\baselineskip 12 pt
\Title{\bf Extracting Parton Densities from Collider Data}
 
\vskip 1.0truein
  
\centerline{\ninerm David A. Kosower}
\baselineskip12truept
\centerline{\nineit Service de Physique Th\'eorique${}^{\dagger}$,
Centre d'Etudes de Saclay}
\centerline{\nineit F-91191 Gif-sur-Yvette cedex, France}
\centerline{\tt kosower@spht.saclay.cea.fr}
      
\vskip 0.5truein
\vglue  0.3cm

\vskip 0.2truein
\baselineskip13truept
\centerline{\bf Abstract}

{\narrower 

Collider data can play an important role in determining the parton distribution
functions of the nucleon.  I present a formalism which makes it possible to
use next-to-leading order calculations in such a determination, while minimizing 
the amount of numerical computation required.

}

\vskip .7 cm 
\centerline{\it Submitted to Nuclear Physics B}

\vfil\vskip .2 cm
\noindent\hrule width 3.6in\hfil\break
${}^{\dagger}$Laboratory of the {\it Direction des Sciences de
la Mati\`ere\/}
of the {\it Commissariat \`a l'Energie Atomique\/} of
France.\hfil\break
       \eject
\footline={\hss\tenrm\folio\hss}

\baselineskip17pt

\noindent
\section{Introduction}
\vskip 10pt

Perturbative quantum chromodynamics gives an excellent description
of short-distance scattering processes at present-day colliders.
The perturbative description relies 
on our ability to compute short-distance matrix
elements in nonabelian gauge theories.
It also relies on our understanding of factorization, which permits
a separation of the process-dependent short-distance aspects from
the universal, process-independent long-distance ones.  The long-distance
parts of scattering processes are captured in the parton distribution
functions of the scattering nucleon(s).  Along with the running coupling
$\alpha_s$, they are the only ingredients needed from outside
perturbation theory for a description (up to subleading power corrections)
of collider scattering processes.  Precise knowledge of parton distributions
is important in the quest for physics beyond the standard model.
At current or planned high-energy colliders new physics must necessarily be detected
against an omnipresent background of QCD or QCD-corrected events.

The momentum evolution of parton distribution functions and of the 
running coupling
is also governed by perturbative equations, so that it is only their
values at a fixed scale which are required inputs from outside 
perturbation theory.  Such input parton distributions may someday be
calculated on the lattice or by other nonperturbative means, but
at present they must be extracted from experiments.
The
modern approaches~[\use\MRS,\use\CTEQ] involve global fits
of next-to-leading order theory
to all available experiments.  The experiments involve different
scale arguments to the distribution functions, but as these are related
by the above-mentioned
 perturbative evolution equation, we can regard the fits as determining
the distributions at a certain fixed scale $Q_0$.  

To date, it is primarily deeply-inelastic scattering data that has been
used in the global fits.  (There is some ad-hoc use of collider data;
for example, MRS~[\use\MRS] make use of certain points from the lepton asymmetry
distribution from CDF and D0, relying on the fact that NLO corrections are
small for these points, and on the availability of an analytic calculation
for this quantity.)  Yet there is a wealth of collider data which may
yield important constraints on the parton densities, in particular the
gluon distribution.  The latter enters into DIS calculations only at higher
order in the coupling, but is a dominant contribution in hadron-hadron
collisions.

The deeply-inelastic structure functions used in the fits can be
described by a convolution of an analytically-known function with the
parton distributions.  This makes their use in a fit computationally 
feasible.  In contrast, for next-to-leading order differential cross
sections at colliders (both hadron-hadron and lepton-hadron),
even the short-distance part must necessarily be calculated numerically.
This is a consequence of the relatively complicated structure of
phase space, once one allows for
arbitrary experimental cuts and jet algorithm.  

In principle, existing NLO jet programs can be used in the global fits.
Were we to try to use them, however, we would have to regenerate the parton
distributions (for example, in the form of numerical tables) at each
iteration of a fit, and then re-run the jet program given the new
values.  This latter part of this procedure would be extremely time-consuming,
and completely impractical for fits requiring more than a handful of
iterations.

Yet much of the calculation of a jet differential cross section --- the
jet algorithm and experimental cuts, the numerical integration
over real emission, the balancing of real and virtual contributions ---
is in fact independent of the precise form of the parton distribution 
functions.  Thus it would be useful to re-organize the calculation
so as to minimize the amount of computational work needed at each
iteration of a fit to experimental data.  

One might imagine pre-generating values for a grid of parameter values,
and interpolating between them to find the best fit, 
but if in a realistic case one has 15 parameters, and wants (say) to
consider ranges discretized into 10 values each, one would have to 
generate $10^{15}$ different points, clearly a hopeless task.  

Fortunately, there is a better way; read on to discover it.
Graudenz et al.\ [\use\Graudenz] have previously
presented an approach to using certain jet distributions
 from lepton-proton scattering.
The approach presented here has certain elements in common with theirs
(such as the use of Mellin transforms), but
 is fully general.  It can be applied to arbitrary
differential distributions in both lepton-hadron and hadron-hadron
scattering.  It is also free of certain theoretical restrictions and
numerical limitations present in their approach.

\def\LIPS{{\rm LIPS}}
\def\kset{\{k_i\}}

\def\qb{{\overline q}}
\def\subfrac#1#2#3#4{%
{#1^{\vphantom{#3}}_{\vphantom{#4}}\over #2}%
\mskip -5mu{\vphantom{#1}^{#3}_{#4}\atop\vphantom{#2}}}
\def\textsubfrac#1#2#3#4{%
{#1^{#3}_{#4}/#2}}

\section{A Prototype}
\tagsection\ToySection
\vskip 10pt

Let us first consider leading-order calculations for a glueball $G$
in a quarkless version of QCD.  This will serve as a warm-up exercise
for the real-world case.

In this case we have only one distribution to consider, $f_{g\leftarrow
 G}(x,Q^2)$.  The total $n$-jet cross section, subject to experimental cuts,
 in glueball-glueball scattering, is given by
$$\eqalign{
\sigma_n &= \int_0^1 \int_0^1 dx_1 dx_2\; 
  \int d\LIPS(x_1 k_G+x_2 k'_{G} \rightarrow \{k_i\}_{i=1}^n) 
\cr &\hskip 10mm\times \vphantom{\int}
     f_{g\leftarrow G}(x_1,\mu^2_F(\kset,x_{1,2}))
     f_{g\leftarrow G}(x_2,\mu^2_F(\kset,x_{1,2}))\,
\cr &\hskip 20mm\times \vphantom{\int}
     \alpha_s^n(\mu^2_R(\kset,x_{1,2}))\hat\sigma(g g\rightarrow \kset)\;
     J_{n\leftarrow n}(\kset)\,,\cr
}\eqn\ToyCrossSection$$
where $\LIPS$ stands for the Lorentz-invariant phase-space measure.  In this
equation, $\hat\sigma$ stands for the usual leading-order partonic
differential cross section with the running coupling $\alpha_s$ set to 1, and
$f_{g\leftarrow G}(x,\mu^2)$ is the gluon distribution inside the glueball.
Note that the $k_i$ are implicitly dependent on $x_1$ and $x_2$ as well.  The
renormalization and factorization scales $\mu_R$ and $\mu_F$ --- typically
something like a jet $E_T$ --- also depend on $x_{1,2}$ and the final-state
momenta (thereby violating, for example, one of the theoretical restrictions
in the work of Graudenz et al.~[\use\Graudenz]).

The jet algorithm is represented by $J_{n\leftarrow n}$, which evaluates to
$1$ if the original $n$-parton configuration yields $n$ jets satisfying the
experimental cuts, and $0$ otherwise.  The precise form of the jet algorithm
is not important for the formalism presented here.

The parton distribution functions satisfy evolution equations, whose solutions
can be written using a {\it universal\/} evolution operator
$E(x,\alpha_s(Q^2),\alpha_0)$.  (The initial coupling $\alpha_0 =
\alpha_s(Q_0^2)$ is one of the parameters we will want to fit.)  Explicit
forms for the evolution operator can be found in the paper of Furmanski and
Petronzio [\use\FP] and elsewhere.  The solutions can be written in the form
$$
f_{g\leftarrow G}(x,Q^2) = E(x,\alpha_s(Q^2),\alpha_0) \otimes 
                             f_{g\leftarrow G}(x,Q_0^2)\,,
\anoneqn$$
where
$$
A(x) \otimes B(x) = \LB A \otimes B\RB(x) \equiv
\int_0^1 dy \int_0^1 dz\; 
\,\delta(x-yz)A(y) B(z) 
\anoneqn$$ 
defines the convolution symbol $\otimes$.

A Mellin transformation turns these convolutions into multiplications,
$$
f_{g\leftarrow G}^z(Q^2) = E^z(\alpha_s(Q^2),\alpha_0) 
                             f_{g\leftarrow G}^z(Q_0^2)\,,
\eqn\ToyMomentEvolution$$
in which
$$
A^z = \int_0^1 dx\; x^{z-1} A(x)
\anoneqn$$
is the Mellin transform of $A$.

To recreate the original function, use the inverse Mellin transform,
$$
A(x) = {1\over2\pi i}\int_{c-i\infty}^{c+i\infty} dz\; x^{-z} A^z\,.
\anoneqn$$
The contour should be chosen to the right of all the singularities
of $A^z$.

We can thus write
$$
f_{g\leftarrow G}(x,Q^2) = 
{1\over 2\pi i}\int_{c-i\infty}^{c+i\infty} dz\; x^{-z} 
E^z(\alpha_s(Q^2),\alpha_0)  f_{g\leftarrow G}^z(Q_0^2)\,.
\anoneqn$$
All the parameters (except $\alpha_0$) that we wish to fit are
contained in $f_{g\leftarrow G}^z(Q_0^2)$.  This latter function
is {\it independent\/}
of all integration variables except $z$, and thus can be pulled out of
the numerical integrations in eqn.~(\use\ToyCrossSection).  The remaining
$z_{1,2}$ contour integrals are to be performed during the fitting
procedure, but this involves only a double sum of the gluon
distribution function multiplied
by precomputed numerical coefficients.  We will be able to make use
of techniques described in ref.~[\use\EfficientEvolution] to find
numerically efficient contours for the $z_i$ integrations.
I postpone the discussion of
the choice of contours, and the method for performing the contour
integrals, to a later section.

Thus at each step of the fitting procedure, we must compute only
$$
-{1\over 4\pi^2} \int_{c-i\infty}^{c+i\infty} dz_1\,
\int_{c-i\infty}^{c+i\infty} dz_2\; 
f_{g\leftarrow G}^{z_1}(Q_0^2)
f_{g\leftarrow G}^{z_2}(Q_0^2) \Sigma^{z_1, z_2}\,,
\anoneqn$$
where $\Sigma^{z_1,z_2}$ are precomputed coefficients given by
$$\eqalign{
\Sigma^{z_1,z_2} &= \int_0^1 \int_0^1 dx_1 dx_2\; 
  \int d\LIPS(x_1 k_G+x_2 k'_{G} \rightarrow \{k_i\}_{i=1}^n) \cr
&\hskip 10mm\times
x_1^{-z_1} x_2^{-z_2} 
E^{z_1}(\alpha_s(\mu^2_F(\kset,x_{1,2})),\alpha_0)
E^{z_2}(\alpha_s(\mu^2_F(\kset,x_{1,2})),\alpha_0)\cr
&\hskip 13mm\times
     \alpha_s^n(\mu^2_R(\kset,x_{1,2}))\hat\sigma(g g\rightarrow \kset)\;
     J_{n\leftarrow n}(\kset)\,.\cr
}\eqn\ToySigmaCoeff$$

Since $\Sigma$ contains all of the short-distance process-specific
dynamics, but is independent of the parton distributions (and indeed,
of the nature of the parent hadron), it is appropriate to call it
a {\it universal cross section\/}, in this case to leading
order in quarkless QCD.  

This procedure works for the parameters in 
$f_{g\leftarrow G}^{z}(Q_0^2)$ because $\Sigma$ is independent of them.
  It does not work for $\alpha_0$, the
remaining fit parameter, because cross sections and distributions depend
on $\alpha_0$ in a (complicated) non-polynomial fashion.  There are
several approaches we can take here.  The simplest is to generate the 
coefficients $\Sigma^{z_1,z_2}$ for a set of $\alpha_0$ around a
``canonical'' value (e.g. $\alpha_s(M_Z^2) = 0.117$), and then fit
using interpolation.  While this approach would be vastly too time-consuming
for a large number of parameters, it is acceptable for a lone parameter.

In general, we don't want to fit total cross sections, but rather
differential distributions.  The above discussions go through 
just as well for the latter.

\section{Leading-Order Fits}
\vskip 10pt

Let us turn next to leading-order fits in the real world, 
specializing to the Tevatron.  The formul\ae\ are
quite similar to those in the previous section, except that we must 
sprinkle a variety of indices in appropriate places.

\def\pb{{\bar p}}
The $n$-jet differential cross section (in the variable $X$) is now given by
$$\eqalign{
{d\sigma_n^{\rm LO}\over dX} &= \sum_{ab} \int_0^1 \int_0^1 dx_1 dx_2\; 
  \int d\LIPS(x_1 k_p+x_2 k_{\pb} \rightarrow \{k_i\}_{i=1}^n) 
\cr &\hskip 10mm\times \vphantom{\int}
     f_{a\leftarrow p}(x_1,\mu^2_F(\kset,x_{1,2}))
     f_{b\leftarrow \pb}(x_2,\mu^2_F(\kset,x_{1,2}))\,
     \delta(X-X(\kset))
\cr &\hskip 20mm\times \vphantom{\int}
     \alpha_s^n(\mu^2_R(\kset,x_{1,2}))
      \hat\sigma^{\rm LO}(a b\rightarrow \kset)\;
     J_{n\leftarrow n}(\kset)\,.\cr
}\eqn\LOCrossSection$$
The partonic cross section $\hat\sigma$ is implicitly summed
over all different possible final states, and the partonic
types $a,b$ are summed over the gluon and
all relevant quark and antiquark distributions.

Eqn.~(\use\ToyMomentEvolution) is replaced by
$$
f_{a\leftarrow p}^z(Q^2) = E^z_{ab}(\alpha_s(Q^2),\alpha_0) 
                             f_{b\leftarrow p}^z(Q_0^2)\,.
\eqn\MomentEvolution$$
The matrix $E^z$ is most easily expressed in a basis of the evolution
eigendistributions $q-\qb$, 
$q+\qb$, along with gluon distribution $g$
and the quark singlet distribution $S$.  The initial parton distributions
are often expressed in this basis as well, but we need
the evolved distributions in the usual flavor basis.  It is therefore
convenient to use $E^z$ in a form where its left index is in the flavor
basis, and its right index is in the eigendistribution basis.

Each step of the fitting procedure involves the computation of
$$
-{1\over 4\pi^2} \int_{c-i\infty}^{c+i\infty} dz_1\,
\int_{c-i\infty}^{c+i\infty} dz_2\; 
f_{a\leftarrow p}^{z_1}(Q_0^2)
f_{b\leftarrow \pb}^{z_2}(Q_0^2) \subfrac{d\Sigma}{dX}{z_1, z_2}{ab}
\eqn\FitStep$$
(with implicit summation over $a,b$), where
$$\eqalign{
\subfrac{d\Sigma}{dX}{z_1,z_2}{ab} &= \int_0^1 \int_0^1 dx_1 dx_2\; 
  \int d\LIPS(x_1 k_p+x_2 k_{\pb} \rightarrow \{k_i\}_{i=1}^n) \cr
&\hskip 10mm\times
x_1^{-z_1} x_2^{-z_2} 
E^{z_1}_{a'a}(\alpha_s(\mu^2_F(\kset,x_{1,2})),\alpha_0)
E^{z_2}_{b'b}(\alpha_s(\mu^2_F(\kset,x_{1,2})),\alpha_0)
\cr&\hskip 13mm\times
     \alpha_s^n(\mu^2_R(\kset,x_{1,2}))
     \hat\sigma^{\rm LO}(a' b'\rightarrow \kset)\;
     J_{n\leftarrow n}(\kset)\, \delta(X-X(\kset))\,.\cr
}\eqn\SigmaCoeff$$
In a numerical computation, the delta function is implemented
by binning in $X$.
 
\section{Next-to-Leading Order Fits}
\tagsection\NextToLeadingOrderFits
\vskip 10pt

\def\smin{s_{\rm min}}
\def\e{\epsilon}
At next-to-leading order, the structure of the cross section is more
complicated; we must combine virtual corrections with real-emission ones.
This is a delicate procedure, because each of these contributions is
independently infrared divergent, and only their sum is well-defined.
Moreover, from a practical point of view, the only practical infrared
regulator is dimensional regularization, which does not mesh naturally
with numerical calculations.  The notion of a parton resolution ---
most simply an invariant mass $\smin$, with a pair of partons $(i,j)$ 
unresolvable if $(k_i+k_j)^2 < \smin$ --- offers a safe passage
through these treacherous divergences and cancellations.  The basic
idea is to evaluate analytically the infrared-divergent contributions
from real emission over the phase space for 
(color-adjacent) {\it unresolved\/} partons,
and to add this contribution to the virtual corrections.  The poles
in the dimensional regulator $\e$ then vanish, and one can take the
four-dimensional limit of all expressions.  The remaining contributions,
over the phase-space for resolved partons, are computed numerically.
Upon adding the two contributions, the dependence on the resolution
parameter disappears (in the limit $\smin\rightarrow 0$) and one
obtains the next-to-leading order differential cross section.  Color
ordering plays an important role because the notion of resolved or
unresolved partons is defined independently for each color permutation.

Several variants of this approach to general processes in collider
physics have been presented in the literature,
in particular the `slicing' [\use\GG,\use\GGK] and 
`subtraction' [\use\FKS,\use\CS] ones.  The details are again not
important for our purposes.  The exposition below assumes the
use of either the pure slicing or a `restricted subtraction' method,
for which a subtraction of singular pieces is performed only inside
the region $s_{\rm sing} < \smin$.  Analogous formul\ae\ can however be
written down for the subtraction method of refs.~[\use\FKS,\use\CS].
I shall indicate the restriction to resolved configurations via
an additional superscript R on, and an additional argument $\smin$ to,
the appropriate leading-order partonic cross-section.  Similarly,
a superscript U will denote the restriction to unresolved configurations.

Thus
\def\ds{\delta\sigma}
\def\dsh{\delta\hat\sigma}
\def\sh{\hat\sigma}
$$\eqalign{
\subfrac{d\ds}{dX}{\rm NLO}{n} &= 
  \subfrac{d\sigma}{dX}{\rm LO}{n} 
  +\subfrac{d\ds}{dX}{\rm NLO,U}{n} 
  +\subfrac{d\ds}{dX}{\rm NLO,R}{n} \,.
}\anoneqn$$
The contribution of configurations with $n+1$ resolved partons is,
$$\eqalign{
\subfrac{d\ds}{dX}{\rm NLO,R}{n} &= \sum_{ab} \int_0^1 \int_0^1 dx_1 dx_2\; 
  \int d\LIPS(x_1 k_p+x_2 k_{\pb} \rightarrow \{k_i\}_{i=1}^{n+1}) 
\cr &\hskip 10mm\times \vphantom{\int}
     f_{a\leftarrow p}(x_1,\mu^2_F(\kset,x_{1,2}))
     f_{b\leftarrow \pb}(x_2,\mu^2_F(\kset,x_{1,2}))\,
     \delta(X-X(\kset))
\cr &\hskip 20mm\times \vphantom{\int}
     \alpha_s^{n+1}(\mu^2_R(\kset,x_{1,2}))
      \sh^{\rm LO,R}(a b\rightarrow \kset;\smin)\;
     J_{n\leftarrow n+1}(\kset)\,.\cr
}\anoneqn$$
In this equation, 
$J_{n\leftarrow n+1}(\kset)$ evaluates to one 
if a given $(n+1)$-parton configuration yields
 $n$ detected jets, and vanishes otherwise.  
It is crucial that both the jet algorithm and the differential cross
section under consideration be infrared-safe, to wit the treatment of
an event must not change under the addition of an arbitrarily soft 
gluon, or the splitting of any parton into two collinear partons.
Otherwise, the details of the jet algorithm are again unimportant
for the formalism presented here.

What is more important for us is the structure
of the unresolved pieces,
$$\subfrac{d\ds}{dX}{\rm NLO,U}{n} = 
\subfrac{d\ds}{dX}{\rm NLO,F}{n}
+\subfrac{d\ds}{dX}{\rm NLO,C}{n}\,.
\anoneqn$$

The first term combines the virtual corrections with the singular
integrals over the unresolved phase space, in a crossing-invariant
fashion.  This term would be present whether or not the initial
state contained colored partons,
$$\eqalign{
\subfrac{d\ds}{dX}{\rm NLO,F}{n} &= \sum_{ab} \int_0^1 \int_0^1 dx_1 dx_2\; 
  \int d\LIPS(x_1 k_p+x_2 k_{\pb} \rightarrow \{k_i\}_{i=1}^{n}) 
\cr &\hskip 10mm\times \vphantom{\int}
     f_{a\leftarrow p}(x_1,\mu^2_F(\kset,x_{1,2}))
     f_{b\leftarrow \pb}(x_2,\mu^2_F(\kset,x_{1,2}))\,
     \delta(X-X(\kset))
\cr &\hskip 20mm\times \vphantom{\int}
     \alpha_s^{n+1}(\mu^2_R(\kset,x_{1,2}))
      \dsh^{\rm NLO}(a b\rightarrow \kset;\smin)\;
     J_{n\leftarrow n}(\kset)\,.\cr
}\anoneqn$$

The second term {\it crosses\/} colored partons from the final state
to the initial state; it is thus absent in $e^+e^-$ scattering.  In our
case, it takes the form
$$\eqalign{
\subfrac{d\ds}{dX}{\rm NLO,C}{n} &= \sum_{ab} \int_0^1 \int_0^1 dx_1 dx_2\; 
  \int d\LIPS(x_1 k_p+x_2 k_{\pb} \rightarrow \{k_i\}_{i=1}^{n}) 
\cr &\hskip 10mm\times \vphantom{\int}
     \LB C_{a\leftarrow p}(x_1,\mu^2_F(\kset,x_{1,2});\smin)
     f_{b\leftarrow \pb}(x_2,\mu^2_F(\kset,x_{1,2}))\RP
 \cr &\hskip 15mm\LP
     + f_{a\leftarrow p}(x_1,\mu^2_F(\kset,x_{1,2}))
     C_{b\leftarrow \pb}(x_2,\mu^2_F(\kset,x_{1,2});\smin)\RB
     \,
     \delta(X-X(\kset))
\cr &\hskip 20mm\times \vphantom{\int}
     \alpha_s^{n+1}(\mu^2_R(\kset,x_{1,2}))
      \hat\sigma^{\rm LO}(a b\rightarrow \kset)\;
     J_{n\leftarrow n}(\kset)\,,\cr
}\anoneqn$$
where $C_{a\leftarrow p}$ is a crossing function as introduced in
ref.~[\use\GGK].  The crossing functions are factorization-scheme dependent; 
they can be expressed in terms of 
scheme-independent functions $A_{a\leftarrow p}$
and scheme-dependent functions $B_{a\leftarrow p}$ as follows,
$$
C_{a\leftarrow p}(x,Q^2) = \left({N\over2\pi}\right)
\left[ A_{a\leftarrow p}(x,Q^2)\ln\left({\smin\over Q^2}\right)  
+ B_{a\leftarrow p}(x,Q^2)\right]\,,
\anoneqn$$
with
$$\eqalign{
A_{a\leftarrow p}(x,Q^2) &= K^A_{a\leftarrow b}(x) \otimes f_{b\leftarrow p}(x,Q^2)\,,\cr
B_{a\leftarrow p}(x,Q^2) &= K^B_{a\leftarrow b}(x) \otimes f_{b\leftarrow p}(x,Q^2)\,.\cr
}\anoneqn$$

Expressions for the kernels $K^{A,B}$ are given in 
ref.~[\use\GGK], and expressions for their Mellin moments $K^{A,z}$ and
$K^{B,z}$ in ref.~[\use\EfficientEvolution].

With these moments, we can define
$$
K^{C,z}_{a\leftarrow b} = 
\left({N\over2\pi}\right)
\left[ K^{A,z}_{a\leftarrow b}\,\ln\left({\smin\over Q^2}\right)  
+ K^{B,z}_{a\leftarrow b}\right]\,;
\anoneqn$$
this allows us
to write 
$$
C_{a\leftarrow p}^z(Q^2) =  K^{C,z}_{a\leftarrow b}\,f_{b\leftarrow p}^z(Q^2)\,,
\anoneqn$$
and then using eqn.~(\use\MomentEvolution),
$$
C_{a\leftarrow p}^z(Q^2) =  K^{C,z}_{a\leftarrow b}\,
E^z_{bc}(\alpha_s(Q^2),\alpha_0) 
                             f_{c\leftarrow p}^z(Q_0^2)\,.
\anoneqn$$

\def\dS{\delta\Sigma}
Each step of the fitting procedure involves the computation of the same
quantity as in eqn.~(\use\FitStep), where now
$$\eqalign{
\subfrac{d\Sigma}{dX}{z_1,z_2}{} &=
\subfrac{d\Sigma}{dX}{{\rm LO:}z_1,z_2}{}
  +\subfrac{d\dS}{dX}{{\rm NLO,U:}z_1,z_2}{}
  +\subfrac{d\dS}{dX}{{\rm NLO,R:}z_1,z_2}{}\cr
 &=  \subfrac{d\Sigma}{dX}{{\rm LO:}z_1,z_2}{}
  +\subfrac{d\dS}{dX}{{\rm NLO,F:}z_1,z_2}{}
  +\subfrac{d\dS}{dX}{{\rm NLO,C:}z_1,z_2}{}
  +\subfrac{d\dS}{dX}{{\rm NLO,R:}z_1,z_2}{}\,.
}\anoneqn$$

The first term, $\textsubfrac{d\Sigma}{dX}{{\rm LO:}z_1,z_2}{}$, is given
by eqn.~(\use\SigmaCoeff); the latter quantities are given by the following equations,
$$\eqalign{
\subfrac{\dS}{dX}{{\rm NLO,R:}z_1,z_2}{ab}
  &= \int_0^1 \int_0^1 dx_1 dx_2\; 
  \int d\LIPS(x_1 k_p+x_2 k_{\pb} \rightarrow \{k_i\}_{i=1}^{n+1})
 \; \delta(X-X(\kset)) \cr
&\hskip 10mm\times
x_1^{-z_1} x_2^{-z_2} 
E^{z_1}_{a'a}(\alpha_s(\mu^2_F(\kset,x_{1,2})),\alpha_0)
E^{z_2}_{b'b}(\alpha_s(\mu^2_F(\kset,x_{1,2})),\alpha_0)
\cr&\hskip 13mm\times
     \alpha_s^{n+1}(\mu^2_R(\kset,x_{1,2}))
     \hat\sigma^{\rm LO,R}(a' b'\rightarrow \kset;\smin)\;
     J_{n\leftarrow n+1}(\kset)\,,\cr
}\eqn\SigmaCoeffR$$

$$\eqalign{
\subfrac{d\dS}{dX}{{\rm NLO,F:}z_1,z_2}{ab} &= \int_0^1 \int_0^1 dx_1 dx_2\; 
  \int d\LIPS(x_1 k_p+x_2 k_{\pb} \rightarrow \{k_i\}_{i=1}^n) 
   \; \delta(X-X(\kset))\cr
&\hskip 10mm\times
x_1^{-z_1} x_2^{-z_2} 
E^{z_1}_{a' a}(\alpha_s(\mu^2_F(\kset,x_{1,2})),\alpha_0)
E^{z_2}_{b' b}(\alpha_s(\mu^2_F(\kset,x_{1,2})),\alpha_0)
\cr&\hskip 13mm\times
     \alpha_s^{n+1}(\mu^2_R(\kset,x_{1,2}))
     \dsh^{\rm NLO}(a' b'\rightarrow \kset;\smin)\;
     J_{n\leftarrow n}(\kset)\,,\cr
}\eqn\SigmaCoeffVF$$

$$\eqalign{
\subfrac{d\dS}{dX}{{\rm NLO,C:}z_1,z_2}{ab}&= \int_0^1 \int_0^1 dx_1 dx_2\; 
  \int d\LIPS(x_1 k_p+x_2 k_{\pb} \rightarrow \{k_i\}_{i=1}^n)  \; \delta(X-X(\kset))\cr
&\hskip 10mm\times
x_1^{-z_1} x_2^{-z_2} 
\LB K^{C,z_1}_{a'\leftarrow \hat a}
E^{z_1}_{\hat a a}(\alpha_s(\mu^2_F(\kset,x_{1,2})),\alpha_0)
  E^{z_2}_{b' b}(\alpha_s(\mu^2_F(\kset,x_{1,2})),\alpha_0)
  \RP\cr &\hskip 15mm\LP
+ E^{z_1}_{a' a}(\alpha_s(\mu^2_F(\kset,x_{1,2})),\alpha_0)
  K^{C,z_2}_{b'\leftarrow \hat b}
  E^{z_2}_{\hat b b}(\alpha_s(\mu^2_F(\kset,x_{1,2})),\alpha_0)\RB
\cr&\hskip 13mm\times
     \alpha_s^{n+1}(\mu^2_R(\kset,x_{1,2}))
     \sh^{\rm LO}(a' b'\rightarrow \kset)\;
     J_{n\leftarrow n}(\kset)\,.\cr
}\eqn\SigmaCoeffVC$$

The calculation of $\textsubfrac{d\Sigma}{dX}{z_1,z_2}{}$ for a given 
$(z_1,z_2)$
amounts to doing the usual next-to-leading order calculation, with the
structure functions $f_{a'\leftarrow p}(x,Q^2)$ replaced by 
$x^{-z} E^z_{a'a}$ ($a$ remaining a free index), and the
crossing functions $C_{a'\leftarrow p}(x,Q^2;\smin)$ replaced
by $x^{-z} K^{C,z}_{a'\leftarrow \hat a} E^z_{\hat a a}$.

\section{Integration Contours}
\tagsection\IntegrationContours
\vskip 10pt

The contour integrals in eqn.~(\use\FitStep) must be performed numerically.
This involves several distinct choices.
First, we must choose an integration contour; then, we must choose a set of
points along the integration contour, or in the
case of hadron-hadron scattering, a set of points in the plane defined by
the product of the two contours.  These points must be chosen in advance,
of course, so that we can evaluate the $\Sigma$ matrices at them.  How
should we make these choices?

The simplest choice is to pick the textbook contour, parallel to the imaginary
axis, but displaced to the right.  This is what was done in the
old papers of Gluck and Reya [\use\GR] on parton evolution, 
and also in the paper by Graudenz et al.~[\use\Graudenz].
The integral evaluated was actually 
$\int_{c-iT}^{c+iT}$, relying on the fall-off of the integrand as 
$T\rightarrow\infty$ to drop the remaining terms.  This finite-interval
integral was evaluated using Gaussian quadrature.   As discussed by
Gluck, Reya, and Vogt~[\use\GRV] and in 
refs.~[\use\Graudenz,\use\EfficientEvolution], however, the integral falls off rather
slowly in this direction, only as a power of the integration variable
$z$.  For parton evolution, the pole structure is such that one can
freely deform the contour into the left-half plane, whereupon the integrand
falls off exponentially, improving the convergence of a numerical evaluation.
Graudenz et al.~[\use\Graudenz] used the Mellin moments of the short-distance
cross section in their formalism; these do not fall off in the left-half
plane, and thus they were not able to deform the contour.  (In contrast,
the numerical Mellin inversions used nowadays in parton evolution programs 
do utilize a rotated contour~[\use\GRV].)

In the formalism presented in previous sections, however, the Mellin moments
of the short-distance cross section appear nowhere.  We may note that
the experimental cuts (on rapidity and jet transverse energy) effectively
impose a minimum on the parton momentum fractions $x_{1,2}$, so that
$\textsubfrac{d\Sigma}{dX}{z_1,z_2}{}$ has no poles in the right-hand plane.
Furthermore, since $0<x_{1,2}<1$, this quantity will also fall off 
exponentially as $z_i\rightarrow \infty$, so long as $\Re z_i < 0$.  Thus
just as in refs.~[\use\GRV,\use\EfficientEvolution],
we can freely shift the contour into the left-half plane, so long as
we stay away from the poles of $f_{a\leftarrow p}^z(Q_0^2)$ along the 
real axis.  Indeed, much of the formalism developed in ref.~[\use\EfficientEvolution]
carries over to the choice of contours in the present paper.

The desired contour would be the contour of steepest descent (which
because of analyticity is also the contour of stationary phase, upon 
which our integrand will be purely real).  For performing the inverse
Mellin transform integrals required for the evolution of parton
distributions, ref.~[\use\EfficientEvolution] shows that it is possible
to construct simple but good analytic approximations to such contours, and
we may hope that those findings carry over to the contours
required for evaluating eqn.~(\use\FitStep).

In any event, we should expect
the contour of steepest descent for the leading-order calculation to
be very close to that for a next-to-leading-order calculation; 
furthermore, we should expect it to not be very sensitive to the 
form of or parameters in the initial parton distributions.  In practice,
we have a reasonable idea of the initial distributions
 (the starting point for the fit, 
say an existing distribution set, will not be radically different
from the best fit), so a good approximation to the contour of steepest
descent we seek will be given by the contour of steepest descent
in eqn.~(\use\FitStep) 
for a leading-order calculation with any existing pdf set.

It will be helpful to examine first the case of lepton-hadron
scattering, which is simpler because there is only one inverse Mellin
transform to compute (and hence a contour in only one variable to choose).
I shall discuss the technically more complicated
hadron-hadron case in the following section.  We can set up the fitting
procedure following the same formalism developed in previous sections.
At leading order, each iteration of 
a fit to a $n$-jet differential cross
section in deeply inelastic scattering (the count excludes the remnant) 
requires computing
$$
{1\over 2\pi i} \int_{c-i\infty}^{c+i\infty} dz\,
f_{a\leftarrow p}^{z}(Q_0^2)\subfrac{d\Sigma}{dX}{z}{a}
\eqn\DISFitStep$$
(with implicit summation over $a$), where
\def\ksete{k'_e,\{k_i\}}
$$\hskip -10pt\eqalign{
\subfrac{d\Sigma}{dX}{z}{a} &= \int_0^1 dx
  \int d\LIPS(x k_p+k_{e} \rightarrow k'_e,\{k_i\}_{i=1}^n) \;
x^{-z} E^{z}_{a'a}(\alpha_s(\mu^2_F(\kset,x)),\alpha_0)
\cr&\hskip 13mm\times
     \alpha_s^{n-1}(\mu^2_R(\ksete,x_{1,2}))
     \hat\sigma^{\rm LO}(e\, a' \rightarrow \ksete)\;
     J_{n\leftarrow n}(\kset)\, \delta(X-X(\ksete))\,.\cr
}\eqn\DISSigmaCoeff$$

In principle, we could let our chosen contour depend on the variable
$X$, but this is both cumbersome and computationally more expensive.
We would rather find a single contour for all values of $X$, so long
as we find a contour (along with 
 a set of point along that contour) that yields an 
accurate evaluation of the
integral for all $X$ where we have data with small statistical and
systematic errors.  We could just take the cross section (the integral
over $X$) but if $X$ depends on the transverse energy $E_T$ ---
that is, if the distribution is differential with respect to the transverse
energy as well as other variables --- then the total cross section
will be dominated by events just above the lower $E_T$ boundary.
This happens because the cross section will be falling rapidly as
a function of $E_T$.  It is probably better, for purposes
of determining the contour, to weight the integral over $X$ by 
a `stratifying' function $S(X)$ which compensates for the rapid
decrease as a function of transverse energy.  Take
$$
F(z) = f_{a\leftarrow p}^{0,z}(Q_0^2)\,
         \int dX\;S(X) \subfrac{d\Sigma}{dX}{z}{a}\,,
\anoneqn$$
where $f_{a\leftarrow p}^{0,z}(Q_0^2)$ denotes the initial 
parton distribution at the start of the fit procedure.
We can now follow the approach outlined in ref.~[\use\EfficientEvolution],
the only change being that derivatives of $F$ must be computed numerically
rather than analytically.  We can, however, write down closed-form
expressions for the required derivatives,
$$\eqalign{
F'(z) &= \int dX\;S(X) \int_0^1 dx\; 
  \int d\LIPS(x k_p+k_{e} \rightarrow k'_e,\{k_i\}_{i=1}^n) \cr
&\hskip 10mm\times x^{-z} 
   \LB -\ln x\, E^{z}_{a'a}(\alpha_s(\mu^2_F(\kset,x)),\alpha_0)
          + {d\over dz}E^{z}_{a'a}(\alpha_s(\mu^2_F(\kset,x)),\alpha_0)\RB
\cr &\hskip 10mm\times
\alpha_s^{n-1}(\mu^2_R(\ksete,x_{1,2}))
     \hat\sigma^{\rm LO}(e\, a' \rightarrow \ksete)\;
     J_{n\leftarrow n}(\kset)\, \delta(X-X(\ksete))\,,\cr
F''(z) &= \int dX\;S(X) \int_0^1 dx\; 
  \int d\LIPS(x k_p+k_{e} \rightarrow k'_e,\{k_i\}_{i=1}^n) \cr
&\hskip 10mm\times x^{-z} 
   \LB \ln^2 x\, E^{z}_{a'a}(\alpha_s(\mu^2_F(\kset,x)),\alpha_0)
      -2\ln x {d\over dz}E^{z}_{a'a}(\alpha_s(\mu^2_F(\kset,x)),\alpha_0)\RP
\cr&\hskip 25mm\LP
      + {d^2\over dz^2}E^{z}_{a'a}(\alpha_s(\mu^2_F(\kset,x)),\alpha_0)
      \RB
\cr &\hskip 10mm\times
\alpha_s^{n-1}(\mu^2_R(\ksete,x_{1,2}))
     \hat\sigma^{\rm LO}(e\, a' \rightarrow \ksete)\;
     J_{n\leftarrow n}(\kset)\, \delta(X-X(\ksete))\,,\cr
F^{(3)}(z) &= \int dX\;S(X) \int_0^1 dx\; 
  \int d\LIPS(x k_p+k_{e} \rightarrow k'_e,\{k_i\}_{i=1}^n) \cr
&\hskip 10mm\times x^{-z} 
   \LB -\ln^3 x\, E^{z}_{a'a}(\alpha_s(\mu^2_F(\kset,x)),\alpha_0)
      +3\ln^2 x {d\over dz}E^{z}_{a'a}(\alpha_s(\mu^2_F(\kset,x)),\alpha_0)\RP
\cr&\hskip 25mm\LP
      -3\ln x {d^2\over dz^2}E^{z}_{a'a}(\alpha_s(\mu^2_F(\kset,x)),\alpha_0)
      + {d^3\over dz^3}E^{z}_{a'a}(\alpha_s(\mu^2_F(\kset,x)),\alpha_0)
      \RB
\cr &\hskip 10mm\times
\alpha_s^{n-1}(\mu^2_R(\ksete,x_{1,2}))
     \hat\sigma^{\rm LO}(e\, a' \rightarrow \ksete)\;
     J_{n\leftarrow n}(\kset)\, \delta(X-X(\ksete))\,.\cr
}\eqn\Derivatives$$

The procedure there can be summarized as follows.  First, find the minimum 
$c_0$ of $F$ along the real axis.  Next define the curve 
$$
z(u) = c_0 + i c_2 \sqrt{u} + c_2^2 c_3 u/2\,,
\anoneqn$$
where 
$$\eqalign{
c_2 &= \sqrt{{2 F(c_0)\over F''(c_0)}}\,,\cr
c_3 &= {F^{(3)}(c_0)\over 3  F''(c_0)}\,.\cr
}\anoneqn$$
To evaluate  eqn.~(\use\DISFitStep), evaluate the transformed version,
$$
{c_2\over 2\pi} \int_0^\infty {du\over\sqrt{u}}e^{-u}\; \Re\LB 
 e^u \L 1-i c_2 c_3\sqrt{u}\R\; 
f_{a\leftarrow p}^{z(u)}(Q_0^2)\subfrac{d\Sigma}{dX}{z(u)}{a}\RB\,,
\eqn\DISFitStepB$$
using a generalized Gauss-Laguerre quadrature,
$$
\int_0^\infty {du\over\sqrt{u}}e^{-u}\; h(u) \simeq \sum_{j=1}^n w_j h(u_j^0)\,,
\eqn\GL$$
where the $u_j^0$ are the zeros of the generalized Laguerre polynomial
$L^{(-1/2)}_n(u)$, and the weights are
given by standard formul\ae~[\use\DR],
$$
w_j = {\Gamma(n+1/2)\over n!\, (n+1)^2} 
   {u_j^0\over \LB L^{(-1/2)}_{n+1}(u_j^0)\RB^2}\,.
\eqn\GLweights$$

The number of points $n$ required for evaluating eqn.~(\use\DISFitStepB) 
to a given desired precision,
using the Gauss-Laguerre quadrature formula~(\use\GL), remains to be
investigated numerically, but experience with parton 
evolution~[\use\EfficientEvolution] suggests
that $n\sim 5$ should be sufficient for $<1\%$ error.

\section{Contours in Hadron-Hadron Collisions}
\tagsection\HadronHadronContours
\vskip 10pt

In hadron-hadron collisions, at each iteration of a fit,
we have not a single contour integral to perform, 
but the double integral~(\use\FitStep).  (I again use the LO
cross section to determine the contours.)  Defining
$$
F(z_1,z_2) = 
f_{a\leftarrow p}^{0,z_1}(Q_0^2)
f_{b\leftarrow \pb}^{0,z_2}(Q_0^2) \int dX\; S(X)
  \subfrac{d\Sigma}{dX}{z_1, z_2}{ab}\,,
\anoneqn$$
we have to find not an approximation to a contour of steepest descent, but 
an approximation to a `surface of steepest descent'.  We can take this
surface to be invariant under conjugation of each $z_i$ separately, and
use this symmetry to rewrite the contour-determining integral,
$$\eqalign{
-{1\over 4\pi^2} &\int_{0}^{c+i\infty} 
\int_{0}^{c+i\infty} \LB 
dz_1\,dz_2\; F(z_1,z_2)
-dz_1\,d\bar z_2\; F(z_1,\bar z_2)
-d\bar z_1\,dz_2\; F(\bar z_1,z_2)
+d\bar z_1\,d\bar z_2\; F(\bar z_1,\bar z_2)\RB\cr
&= -{1\over 2\pi^2} \int_{0}^{c+i\infty} 
\int_{0}^{c+i\infty} \Re\LB 
dz_1\,dz_2\; F(z_1,z_2)
-dz_1\,d\bar z_2\; F(z_1,\bar z_2)\RB\cr
}\eqn\FitStepB$$
It is {\it not\/} necessarily symmetric in $z_1\leftrightarrow z_2$, 
because the beams, detector, or observable (such as a 
parity-violating one) may not satisfy the 
required $x_1\leftrightarrow x_2$ symmetry.

\def\cz{(c_1,c_2)}
The poles in the $z_i$ lie along the negative real axis, so we can
freely deform the contours into the left-half plane.  Now just follow
the procedure used in the single-contour case.
First find the minimum of
$F(z_1,z_2)$ for real $z_{1,2}$, and label its coordinates $\cz$.
Parametrize the surface using two variables $t_{1,2}$,
$$\eqalign{
z_1 &= c_{1} + i t_1 + a_{11} t_1^2 + a_{12} t_1 t_2\,,\cr
z_2 &= c_{2} + i t_2 + a_{22} t_2^2 + a_{21} t_1 t_2\,,\cr
}\eqn\ChangeA$$

Expand the equation $\Im F(z_1(t_1,t_2),z_2(t_1,t_2))=0$ to obtain
$$\eqalign{
a_{11} &= {F^{(3,0)}\over
           6 F^{(2,0)}}\cr
a_{12} &= -{F^{(0,3)}\over 6 F^{(0,2)}}
          +{1\over 6 J}\LB 
          {F^{(0,3)} F^{(2,0)}}
          -3 {F^{(1,2)} F^{(1,1)}}
          +3 {F^{(2,1)} F^{(0,2)}}
          -{F^{(3,0)} F^{(0,2)} F^{(1,1)}\over F^{(2,0)}}\RB\cr
a_{21} &= -{F^{(3,0)}\over 6 F^{(2,0)}}
          +{1\over 6 J}\LB
          {F^{(3,0)} F^{(0,2)}}
          -3{F^{(2,1)} F^{(1,1)}}
          +3{F^{(1,2)} F^{(2,0)}}
          -{F^{(0,3)} F^{(2,0)} F^{(1,1)}\over F^{(0,2)}}\RB\cr
a_{22} &= {F^{(0,3)}\over 6 F^{(0,2)}}\cr
}\anoneqn$$
where I use the abbreviated notation
$$\eqalign{
F^{(j_1,j_2)} &= \LP
\LB{\partial^{j_1+j_2} F(z_1,z_2)\over\partial z_1^{j_1}\partial z_2^{j_2}}\RB
 \RV_{(z_1,z_2)=\cz}\cr
J &= \det\L \matrix{ F^{(2,0)} & F^{(1,1)}\cr
                     F^{(1,1)} & F^{(0,2)}\cr}\R\cr
}\anoneqn$$

We expect the function to go like $F\cz e^{-g(t_1,t_2)}$, with
$$
g(t_1,t_2) = {F^{(2,0)}\over 2 F\cz} t_1^2
+{F^{(1,1)}\over F\cz} t_1 t_2
+{F^{(2,0)}\over 2 F\cz} t_1^2
\anoneqn$$
which suggests the change of variables
$$\eqalign{
t_1 &= \sqrt{{F\cz(\Delta+\Lambda)\over \lambda_{+}\Delta}} \sqrt{u_1}
+\sqrt{{F\cz(\Delta-\Lambda)\over \lambda_{-}\Delta}} \sqrt{u_2}\cr
t_2 &= -\sqrt{{F\cz(\Delta-\Lambda)\over \lambda_{+}\Delta}} \sqrt{u_1}
+\sqrt{{F\cz(\Delta+\Lambda)\over \lambda_{-}\Delta}} \sqrt{u_2}\cr
}\eqn\ChangeB$$
where
$$
\Delta = F^{(2,0)}- F^{(0,2)}\,,
\anoneqn$$
$$
\lambda_{\pm} = {1\over2}\LB F^{(2,0)} + F^{(0,2)}
        \pm \sqrt{\L F^{(2,0)} - F^{(0,2)}\R^2+4 \L F^{(1,1)}\R^2}\RB
\anoneqn$$
are the eigenvalues of the Hessian matrix at $\cz$, and
$$
\Lambda = \lambda_{+}-\lambda_{-}\,.
\anoneqn$$

Using the changes of variables~(\use\ChangeA,\use\ChangeB), we can
rewrite the integral~(\use\FitStep) to be performed at each iteration
of a fit as follows,
$$\eqalign{
{1\over 2\pi^2} &{F\cz\over \sqrt{J}}\int_0^\infty \int_0^\infty du_1 du_2\;
{e^{-u_1-u_2}\over\sqrt{u_1 u_2}}\,h(u_1,u_2)\,,
}\anoneqn$$
with
$$\eqalign{
h(u_1,u_2) &= 
\Re\LB e^{u_1+u_2}\,
\vphantom{ \subfrac{d\Sigma}{dX}{z_1(u_i), z_2(u_i)}{ab} }
\bigl(  1 - i (2 a_{11} + a_{21}) t_1(u_i) 
   - i (a_{12} + 2 a_{22}) t_2(u_i)
\RP\cr &\hskip 25mm\LP
   - 2 a_{11} a_{21} t_1^2(u_i) 
   - 4 a_{11} a_{22} t_1(u_i) t_2(u_i) - 2 a_{12} a_{22} t_2^2(u_i) \bigr)
\RP\cr &\hskip 20mm\LP\times
f_{a\leftarrow p}^{z_1(u_i)}(Q_0^2)
f_{b\leftarrow \pb}^{z_2(u_i)}(Q_0^2) \subfrac{d\Sigma}{dX}{z_1(u_i), z_2(u_i)}{ab}
\RB\,.}\eqn\FinalSummand$$
($z_{1,2}$ and $t_{1,2}$
are functions of $u_{1,2}$ via eqns.~(\use\ChangeA,\use\ChangeB))

We will again use the generalized Gauss-Laguerre quadrature formula,
here in each of the variables $u_i$.  This would ordinarily lead to using
$n^2$ points having chosen the $n$-point formula; we may note, however,
that points at the far end of the square away from the origin will give
a negligible contribution, so that we can restrict the sum to points
$(u_i^0,u_j^0)$ with $u_i^0+u_j^0\leq u_1^0+u_n^0$,
$$
\int_0^\infty {du_1 du_2\over\sqrt{u_1 u_2}}e^{-u_1-u_2}\; h(u_1,u_2) 
   \simeq \sum_{j_1,j_2=1}^n w_{j_1} w_{j_2} h(u_{j_1}^0,u_{j_2}^0)
   \simeq \sum_{j_1,j_2=1\atop u_{j_1}^0+u_{j_2}^0\leq u_1^0+u_n^0}^n 
     w_{j_1} w_{j_2} h(u_{j_1}^0,u_{j_2}^0)
\,.
\eqn\FinalSum$$ 
(The $u_j^0$ are again the zeros of the generalized Laguerre polynomial
$L^{(-1/2)}_n(u)$, and the weights 
$w_j$ are given by eqn.~(\use\GLweights).)

 The number of points needed remains
to be investigated numerically.  Formul\ae\ for the various derivatives
used above can be written for computer evaluation in the same fashion as
in the case of deeply inelastic scattering, eqn.~(\use\Derivatives),
$$\eqalign{
{\partial^{j_1+j_2} F\over\partial^{j_1} z_1 \partial^{j_2} z_2} (z_1,z_2) &= 
f_{a\leftarrow p}^{0,z_1}(Q_0^2)
f_{b\leftarrow \pb}^{0,z_2}(Q_0^2) \int dX\; S(X)
    \int_0^1 \int_0^1 dx_1 dx_2\; 
  \int d\LIPS(x_1 k_p+x_2 k_{\pb} \rightarrow \{k_i\}_{i=1}^n) \cr
&\hskip 10mm\times
\LP{d^{j_1}\over dz^{j_1}}\RV_{z=z_1} 
\LB  x^{-z} E^{z}_{a'a}(\alpha_s(\mu^2_F(\kset,x_{1,2})),\alpha_0)\RB
\cr&\hskip 13mm\times
\LP{d^{j_2}\over dz^{j_2}}\RV_{z=z_2} 
\LB  x^{-z} E^{z}_{a'a}(\alpha_s(\mu^2_F(\kset,x_{1,2})),\alpha_0)\RB
\cr&\hskip 13mm\times
     \alpha_s^n(\mu^2_R(\kset,x_{1,2}))
     \hat\sigma^{\rm LO}(a' b'\rightarrow \kset)\;
     J_{n\leftarrow n}(\kset)\, \delta(X-X(\kset))\cr
}\anoneqn$$
where $f_{a\leftarrow p}^{0,z}(Q_0^2)$ denotes (as in 
section~\use\IntegrationContours) the initial 
parton distribution at the start of the fit procedure.

In a practical application of the formalism presented here, one would
proceed as follows.  One would first determine the contours appropriate
to each distribution one wanted to use in (say) fitting the parton
distribution functions of the proton.  (It is plausible that different
distributions could make use of a common contour, but this is not
guaranteed.)  One would then determine the required number of points
in the inversion procedure.  These computations would be done using
leading-order matrix elements, as dicussed above.  Having chosen
the points $z_{1,2}$, one would then perform a next-to-leading order
computation of the quantities $d\Sigma_{ab}^{\rm LO; NLO, F; NLO, C; NLO, R}/dX$
defined in section~\use\NextToLeadingOrderFits.  One typically
uses functional forms for the initial parton distributions whose 
Mellin transforms are known analytically as a function of the parameters
and of the Mellin variable $z$.  

One would seek to minimize the $\chi^2$ of the fit to experimental data by
an iterative procedure.
The Mellin transforms of the initial parton densities
may be evaluated numerically
for a given choice of the parameters.
The quantity $h$ defined in eqn.~(\use\FinalSummand) is then just a
numerical inner product, with the `metric' given by the $d\Sigma_{ab}/dX$
matrices evaluated previously.
At each iteration of such a procedure, one would use eqn.~(\use\FinalSum)
to compute the desired observable.  Only the numerical values of
the Mellin transforms of the initial parton densities would need to be
evaluated anew at each iteration of the fit.

\section{Conclusions}
\vskip 10pt

Jet data collected at the Tevatron and at HERA can play an important role
in determining the parton distribution functions of the nucleon.  To minimize
renormalization-scale uncertainties, and to take full advantage of the
data, they should be fitted using next-to-leading order (or higher order) theory.
The formalism presented in this paper makes it computationally practical
to fit to distributions culled from jet data, using 
a modern next-to-leading calculation.  It reorganizes the calculation
so that most of the time-consuming computations in an NLO program are done only
once, and that each iteration of a fit involves only the recomputation of the
Mellin transform of the initial parton densities at a few points, and a weighted
sum over those points.  The number of points required can also be minimized via
choice of a quadratic contour, using the same approach detailed in 
ref.~[\use\EfficientEvolution].  A similar approach could be used to reorganize
the calculation of next-to-leading order corrections to allow the extraction
of parton-to-hadron fragmentation functions from collider data.

\listrefs
\bye